# ArchiTone: A LEGO-Inspired Gamified System for Visualized Music Education

JIAXING YU, Zhejiang University, China
TIEYAO ZHANG, Zhejiang University, China
SONGRUOYAO WU, Zhejiang University, China
XINDA WU, Zhejiang University, China
TINGXIAO WU, Zhejiang University, China
YANJUN CHEN, Zhejiang JadeFoci Techonology Co. LTD, China
KEJUN ZHANG*, Zhejiang University, China

Participation in music activities has many benefits, but often requires music theory knowledge and aural skills, which can be challenging for beginners. To help them engage more easily, it's crucial to adopt teaching strategies that lower these barriers. Informed by formative investigation and inspired by LEGO, we introduce ArchiTone, a gamified system that employs constructivism by visualizing music theory concepts as musical blocks and buildings for music education. This system includes two modes: Learning Mode, which involves recognizing and learning common musical blocks through familiar musical works; Creation Mode, which allows learners to freely create and combine musical blocks to produce new musical works. User studies demonstrate that our gamified system is not only more engaging than traditional music education methods but also more effective in helping learners understand abstract music theory and apply it to music praxis. Additionally, learners demonstrate superior performance on music theory tasks after using ArchiTone.

# 1 INTRODUCTION

In recent years, music education has garnered increasing attention across all ages [48, 66]. This growing interest largely stems from the positive effects that successful participation in musical activities has on participants [7, 67], such as reducing stress [17, 71], enhancing emotional regulation [11, 44, 63], and improving social skills [6, 57]. As a discipline focused on the foundational elements of music and their organization, music theory plays a crucial role in music education. The study of music theory enables learners to grasp the principles of musical composition and develop an aesthetic appreciation for musical works. Additionally, learners with a solid foundation in music theory can better participate in musical activities, which enhances their sense of social inclusion [68], thus achieving the core goals of music education.

---

*Corresponding author.

However, learning music theory is challenging. It requires learners to 1) grasp abstract and intricate musical concepts [32], 2) recognize these concepts in musical sounds and apply them flexibly in music praxis. Mastery of aural skills, knowledge, or techniques alone is insufficient to become a true musician [10]. This process demands substantial time and effort for learners to achieve desired outcomes which often discourages them [20]. Meanwhile, traditional music education methods face significant limitations due to their lack of engaging learning strategies and their inability to integrate theory with practice [51, 53]. While music education nowadays has increasingly acknowledged these challenges [18, 19], addressing them effectively often requires meticulous course design and guidance from teachers, thereby placing greater demands on their professional expertise [23].

If learners can actively engage in the process of constructing music theory knowledge and exploring it based on their own experiences, they will gain a deeper understanding of music theory concepts and apply them more effectively in music praxis. Then learning music theory will no longer be a barrier to music education. This idea aligns with the principles of constructivism. The LEGO learning system, as a representative of constructivism, is recognized for its intuitiveness, inclusivity, and adaptability, making it suitable for learners of all ages. It has demonstrated clear advantages over traditional education methods [1, 26, 36]. Hence, we propose that the LEGO learning system could provide a novel approach to music education. By visualizing key music theory concepts—such as chords and harmony structures—as musical blocks similar to LEGO building blocks, and allowing learners to combine these blocks in various ways, it facilitates both the study of music theory and the process of music praxis. This approach simplifies the complexities of music theory, offering a more intuitive and interactive learning experience that helps learners overcome challenges, leading to a deeper understanding and more effective application of music theory.

Furthermore, with the rapid advancement of information technology, mobile education applications are gaining increasing attention in music education and offering significant benefits [56, 65]. By designing a gamified interface that runs on mobile devices (smartphones and tablets), we can promote a style of music education that reaches a broader audience. This approach will facilitate access to music education, enhance learners' intrinsic motivation [4, 14], enable effective learning anytime and anywhere [47], and support the development of creativity [73].

Building on related work and our own formative investigation, we introduce ArchiTone, a gamified music education system designed for mobile devices. The core concept of ArchiTone is to blend music theory with game mechanics, and to simplify user comprehension through the visualization of concepts and rules. The system employs constructivism by visualizing chords and harmony structures, the fundamentals of music theory, as musical blocks. And users are guided to combine these musical blocks based on music theory principles to achieve music education. The system includes two modes: Learning Mode, which helps users recognize and learn common musical blocks through well-known musical works, and Creation Mode, which allows learners to freely create and combine musical blocks to create new musical works.

In this paper we present the following main contributions:

- A formative investigation of the needs of music educators and learners suggests that music education tools should visualize abstract music theory concepts into easily understandable elements, encourage learners to interact repeatedly and actively, and apply the acquired knowledge in practice to achieve the goals of music education.

- ArchiTone, a LEGO-inspired gamified system designed for mobile devices that applies the principles of constructivism to achieve both the learning of music theory and its application in music praxis.
- User studies demonstrate the usability and effectiveness of ArchiTone in music education by evaluating both user-reported experience and objective measures of music theory.

## 2 RELATED WORK

ArchiTone adopts the principles of constructivism, specifically the LEGO education system, visualizes key music theory concepts, and develops a gamified system for music education. In this section, we describe related work on constructivism and the LEGO education system in education, music visualization, and music education systems.

### 2.1 Constructivism in Education

Constructivism, as a learning theory, has significantly influenced educational practices by emphasizing the role of learners as active participants in knowledge construction. Piaget and Vygotsky laid the foundation for constructivism in education, highlighting the importance of experiential learning and social interaction [2]. Their work has spurred research in instructional design that supports learner-centered environments, where learners construct understanding through problem-solving and collaboration [28]. Recent studies [21, 22, 37, 42, 55] have expanded on these principles, integrating technology into constructivism to create dynamic, interactive learning experiences. The LEGO education system exemplifies this integration by providing hands-on learning environments that foster creativity and critical thinking, aligning with the principles of constructivism.

*2.1.1 LEGO Education System.* As a representative of constructivism, the LEGO education system helps learners explore abstract concepts through tangible shapes and patterns, ensuring their active engagement, while also developing a range of critical skills [45]. Its effectiveness in education has been well established [1, 12, 27, 43, 59]. [1] studies the impact of the system on elementary students' math performance and proposes an innovative tool, MoreToMath. The study finds that MoreToMath can attract the interest of students with different mathematical ability levels and improve their math performance. [12] investigates the use of the LEGO education system in early childhood and preschool education programs, showing that the system effectively meets educational objectives and promotes the development of children in many aspects. Additionally, researchers have adapted the LEGO education system in many creative ways. [27] develops an immersive music composition course for young children, where they utilize LEGO blocks to create three-dimensional representations of musical works. This approach enables young children to achieve advanced music education standards and produce unique musical works. Inspired by constructivism and the LEGO education system, we propose an accessible and engaging method that allows learners to easily grasp music theory and apply their knowledge in music praxis.

### 2.2 Music Visualization

Music visualization involves utilizing computer graphics to depict musical performances on a static or dynamic picture, aiming to enhance audience's understanding and emotional connection to the music through visual representation [25]. It can be categorized into two types: augmented scores and performance visualization. Augmented scores enhance traditional music sheets by integrating visual elements [58, 61] such as colors and shapes to make the analysis and learning of music more intuitive. Performance visualization [24] emphasizes the real-time analysis of music, uncovering

Conference acronym 'XX, June 03–05, 2018, Woodstock, NY                                                                                      Trovato et al.its intrinsic structure and emotional expression through the visualization of dynamic features such as volume, pitch, and melody.

Among them, interactive music visualization systems have gained attention due to their unique advantages [16, 34, 35]. VisualHarmony [39] is an innovative tool designed to assist classical music enthusiasts and professionals in composition. It uses a visual interface to display the harmony structure of music in real time, helping users understand and avoid harmonic errors during the creation process. MoshViz [8] presents a visualization framework designed for in-depth music analysis. By integrating the detail view and the overview view, users can quickly explore and comprehend the structure of musical compositions. The framework leverages interactive mechanisms such as linking and brushing strategies and parameter adjustments, offering a flexible platform for music exploration. Users can seamlessly switch between different levels of views, with changes in color and size helping to identify musical patterns, thereby achieving a comprehensive insight into the musical work.

Although existing interactive music visualization methods are functionally advanced and highly specialized, they typically require users to have a foundational knowledge of music theory. To address this challenge, we design a new music visualization tool that is accessible to learners of all ages, which helps them learn music theory quickly through an intuitive interface and apply it to musical activities.

### 2.3 Music Education System

In recent years, advancements in technology and educational theory have given rise to many music learning tools. These tools increase learners' efficiency and engagement, while also reducing the barriers to music education. Gamification and cognitive learning theory are two strategies that are widely used in music education systems.

*2.3.1 Gamification.* Gamification, as an effective educational method [3, 9, 30, 31], skillfully balances enjoyment with educational goals. [15] develops a music education game that enables players to use a game controller as an instrument, stimulating players' curiosity and creativity. Troubadour [46] is a gamified learning platform designed to improve students' sight-singing and ear-training skills through personalized and adaptive music theory exercises. After using Troubadour, learners achieve better performance on music theory exams. The study also identifies gamified elements and a user-friendly interface as crucial factors in increasing student motivation and overall learning experience. Bean Academy [33] is a music composition game that employs the Vocal Query Transcription (VQT) model. This model allows users to input music by humming, which is then converted into digital sound. By minimizing the need for advanced music theory, the game enables beginners to get started quickly, thereby increasing their satisfaction in learning music composition. ChordAR [38] combines augmented reality (AR) with artificial intelligence (AI) to provide immersive music theory education for preschool children. By utilizing multi-modal input and cloud-based AI training, the system effectively monitors physical interactions and gives children a better understanding of chords and melody perception.

*2.3.2 Cognitive Learning Theory.* Cognitive Load Theory (CLT) highlights that, due to the limited capacity of human working memory [52, 60], education system designers should minimize cognitive load on students. Brain Automated Chorales (BACh) [72] employs near-infrared spectroscopy to assess users' cognitive workload in adaptive learning systems. This technology allows for automatic adjustment of learning difficulty, facilitating piano instruction of Bach's chorales within optimal cognitive load levels. Feedback from participants suggests that they find BACh to be more effective for learning and value the well-timed adjustments in difficulty levels.

The work described above targets specific aspects of music education and focus on reducing learning barriers or increasing engagement. However, these approaches have not yet provided a comprehensive learning experience that integrates music theory with practice. Balancing theory and practice is crucial, as it allows learners to explore new possibilities through hands-on experience and to build upon existing knowledge more effectively. In contrast, a disconnect between theory and practice can hinder learners' musical development [69].

## 3 FORMATIVE INVESTIGATION

As detailed above, there are already a variety of systems, tools, and approaches designed to support music education. However, they often struggle to address the needs of aural perception, conceptual understanding, and music praxis simultaneously. We engaged in a formative investigation to 1) understand the disconnect between existing solutions and educator and learner needs, and 2) provide design decisions that target the correct challenges.

### 3.1 Method

We carried out in-depth interviews with three music educators (2 female, 1 male) and four music learners (ages 21–32; 1 female, 3 male). To ensure the diversity and inclusivity of our investigation, we recruited educators working in a variety of settings, including formal classrooms and after-school programs, as well as both full-time and part-time learners. The interviews were conducted in a semi-structured format, focusing on three key topics: 1) experiences in teaching or learning music, 2) daily learning activities, and 3) the challenges and motivations for such teaching and learning. The interviews lasted 20 to 40 minutes with an average duration of 29.6 minutes ($M = 29.6$) for educators and 10 to 18 minutes with an average duration of 16.2 minutes ($M = 16.2$) for students. Throughout the interview, the interviewer meticulously documented participants' responses and posed targeted follow-up questions based on the responses. Furthermore, the interview was audio-recorded and subsequently transcribed for later comprehensive analysis.

### 3.2 Key Findings & Design Implications

From the aforementioned process, we identified three key findings and derived corresponding design implications. These findings and implications highlight the challenges encountered by both educators and learners in music education, and outline the strategies we implemented to address these challenges.

*3.2.1 Abstractness of Music Theory.* For most music learners, our observations revealed that music education is not accessible, primarily because they struggle to grasp abstract music theory, which has also been discussed in previous studies [32]. Additionally, we found that the difficulty in understanding music theory lies in the need to map abstract concepts to concrete musical sounds. Learners usually lack the ability to establish connections between theoretical concepts and their aural representations. To tackle the challenge, we aim to 1) make abstract concepts visual and interactive, and 2) to promote the integration and application of knowledge through analogies.

*3.2.2 Repetition in Music Training.* We also identified that even if learners can understand abstract music theory, the repetitive training to establish connections between concepts and musical sounds often lacks interest and can easily lead to negative experiences, making it difficult for continuous learning. Repeated engagement in music training is crucial, as it not only reinforces the connections but also helps build proficiency and confidence. To maintain engagement, both educators and learners agreed to incorporate varied and interactive contents that keep the learning process dynamic and motivating. Based on these comments, we aim to promote engagement by 1) leveraging gamification to enhance learner interaction, and 2) providing timely feedback during interactions.

*3.2.3 Purposefulness of Music Education.* Finally, as repeatedly mentioned by educators and learners, music education is highly purposeful. Learners tend to focus on acquiring musical knowledge that is directly related to their praxis, while showing a negative attitude toward unrelated content. Although this musical knowledge may not significantly enhance learners' music skills or techniques, it can deepen their understanding of music and help them form a complete music cognitive system. Moreover, this improvement in musical understanding is not immediately apparent in the short term; learners need continuous music praxis to gradually perceive and appreciate this progress. Building on these key findings, we aim to 1) achieve sustainable and incremental learning through constructivism, and 2) enable learners to efficiently apply the acquired knowledge in practice.

## 4 SYSTEM DESIGN

Based on the above insights, we design and develop ArchiTone, a gamified system for music education working on mobile devices that incorporates the principles of constructivism and the LEGO education system to facilitate both the learning of music theory and the creation of musical works. Fig. 1 illustrates the process of ArchiTone. In the subsequent sections, we will detail the design objectives, our music visualization and interaction design, and the user flows of music theory learning in Learning Mode and music composition practice in Creation Mode.

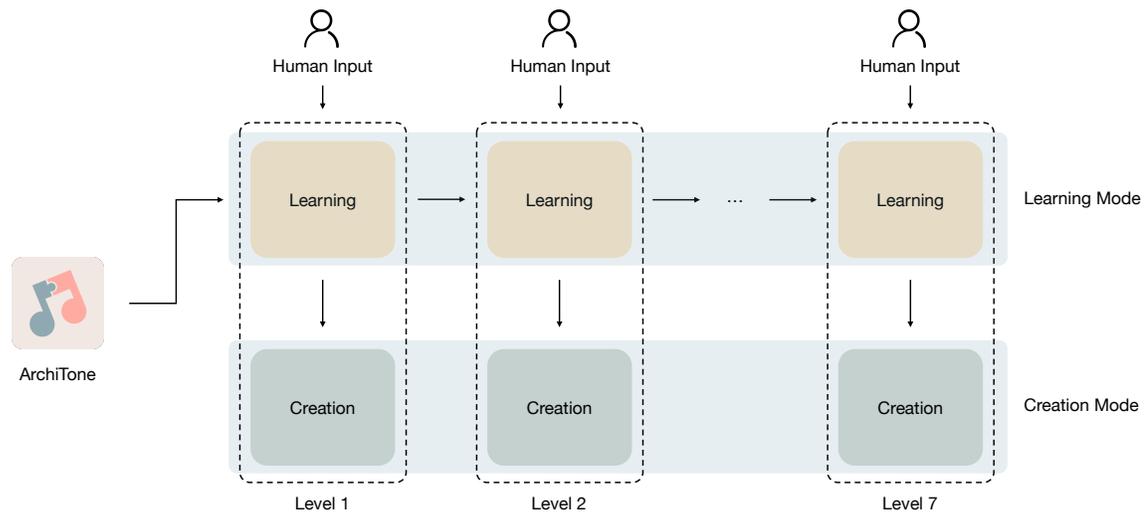

Fig. 1. Overview of ArchiTone procedure. ArchiTone consists of seven levels, each containing two parts: music theory learning in Learning Mode and music composition practice in Creation Mode.

### 4.1 Design Objectives

In alignment with our previously stated goals to address the abstractness of music theory, we develop a visualization scheme for music concepts, particularly targeting chords and harmony structures, which are both fundamental and challenging elements in learning music theory. Specifically, we visualize chords as musical blocks that can be connected to each other and harmony structures as musical buildings, similar to blocks and buildings in LEGO. We initially apply this visualization scheme to musical works that are familiar to learners in Learning Mode of the system, which helps them quickly grasp both the visualization approach and the interactive techniques.

To support our engagement and purposefulness goals, we leverage constructivism to design a gamified system that operates on mobile devices and offers varying levels of difficulty, starting from basic levels to advanced levels, thereby facilitating sustainable and incremental learning. The system allows learners to freely explore combinations of different musical blocks in Creation Mode and enables them to efficiently apply the acquired knowledge in practice. Additionally, throughout the exploration process, our system provides real-time feedback to enhance immersion and engagement in the interaction.

### 4.2 Music Visualization and Interaction Design

For chords and harmony structures in music theory, we utilize visualization and interaction design to intuitively display their internal relationships, providing learners with a more coherent system for music understanding and composition.

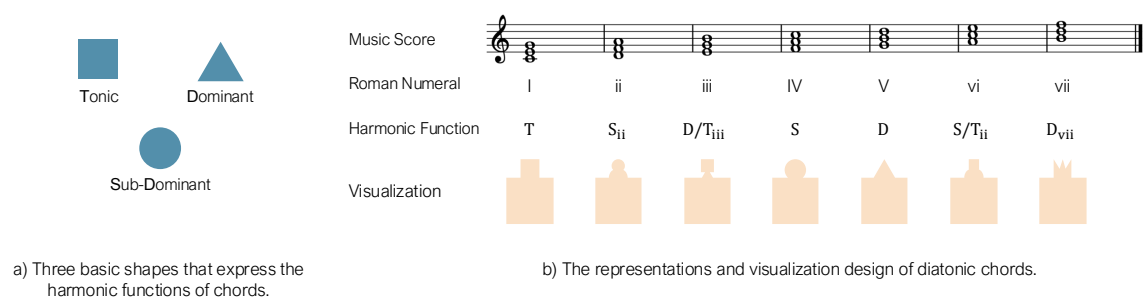

a) Three basic shapes that express the harmonic functions of chords.

b) The representations and visualization design of diatonic chords.

Fig. 2. The visualization design of harmonic functions and chords. a) Three basic shapes for representing harmonic functions of chords: square for tonic chords, triangle for dominant chords, and circle for sub-dominant. b) Different representations of diatonic chords and the visualization design according to their harmonic functions.

*4.2.1 Visualization Design for Chords.* Chords are defined as a harmonic set of three or more notes that are played simultaneously or in close succession to produce a cohesive sound, and can be categorized and labeled according to their harmonic functions. We begin by expressing three harmonic functions using three common shapes: square for tonic chords, triangle for dominant chords, and circle for sub-dominant, as shown in Fig. 2(a). The square symbolizes stability and is used to denote tonic chords, embodying a sense of solidity within the music. The triangle represents dominant chords, with its sharp angles conveying a sense of tension and instability. The circle is designated for sub-dominant chords, with its gentle curves reflecting a smooth and mellow tonal quality. These three shapes effectively illustrate the aural experiences associated with chords of different harmonic functions: stable, tense, and smooth. Then, we design seven different symbols for the seven diatonic chords (I, ii, iii, IV, V, vi and vii) in a key, according to their specific harmonic functions, as shown in Fig. 2(b). Specifically, for the I, IV and V chords, they only have a single harmonic function and are represented directly by the corresponding shapes. For the ii and vii chords, they have stronger harmonic functions than IV and V and are indicated by a combination of multiple identical shapes. For the iii and vi chords, they exhibit two harmonic functions and are represented using overlapping patterns of corresponding symbols.

*4.2.2 Visualization Design for Harmony Structures.* Harmony structure refers to the overall organization of chords in a musical composition, typically manifested through chord progressions. Based on Schenker's harmonic analysis theory, we design three types of musical block combinations, with each type representing a kind of harmony structure, as illustrated in Fig. 3. Specifically, horizontally arranged musical blocks denote the natural progression of chords within the harmony structure, whereas blocks situated above the harmony structure represent prolongation. For the prolongation, there are two primary chord progression types: neighbor progression and passing progression. Neighbor progression is symbolized by a vault building, with the roof being detailed according to the number of chords. It visually emphasizes the transience of neighbor tones and their subtle influence on the harmony structure. Passing progression is embodied by a building composed of musical blocks stacked one above the other, which conveys the fluidity of passing tones and their ongoing influence on the chord progression. The design for harmony structures provides a visual approach that simplifies the complexity of chord progressions, making it more understandable and easier to learn.

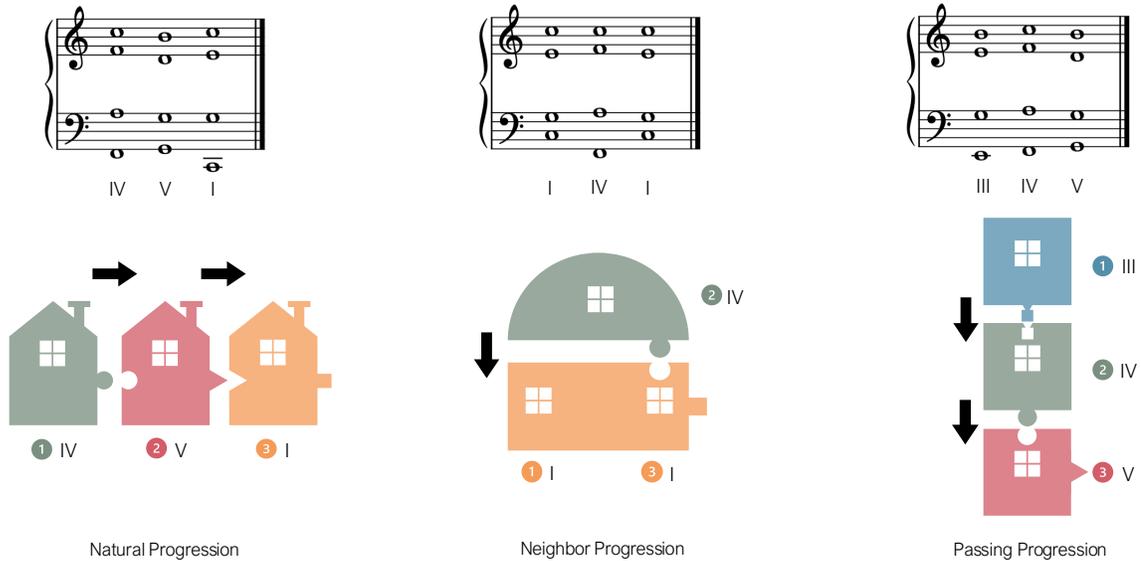

Fig. 3. The visualization design for three harmony structures: natural progression, neighbor progression and passing progression.

*4.2.3 Interaction Design.* Alongside the visualization design, we also conduct a series of interaction designs to enable learners to use the system more intuitively and enhance the overall engagement. Firstly, we implement a mortise and tenon structure, inspired by traditional Chinese architecture. This design connects the musical blocks through the precise interlocking of the tenon (the projecting part) and the mortise (the recessed part), ensuring a stable and secure connection. In our system, the tenon-and-mortise relationship is used to represent the connections between chords. If the tenon of one chord does not properly align with the mortise of the subsequent chord, it indicates that the two chords cannot be connected.

Additionally, we integrate multiple real-time feedback mechanisms to enhance learners' immersion and engagement. 1) Position detection: the system detects the positions of different musical blocks in real time for immediate feedback. 2) Attraction and repulsion: when two compatible blocks come within a certain distance, they will automatically snap together with an animation and a "click" sound. If the blocks are not compatible, they will repel each other. 3) Hints: the system features multiple interfaces with appropriate hints. During learning or creation, real-time notifications are provided if actions significantly deviate from the system's expectations.

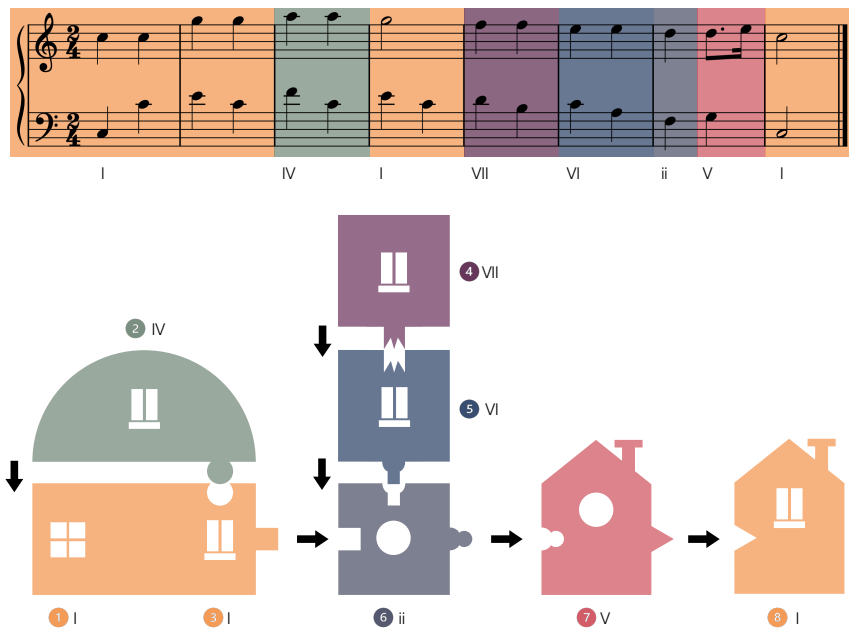

Fig. 4. Example of a music piece from 'Twinkle, Twinkle, Little Star' represented using our visualization design.

### 4.3 User Flows

Our system consists of two main user flows: music theory learning through the Learning Mode and music composition practice through the Creation Mode.

*4.3.1 Music Theory Learning.* The music theory learning flow is the central part of ArchiTone, designed to progressively introduce learners to chords and harmony structures while building connections between theoretical knowledge and corresponding musical sounds. In Learning Mode, there are seven levels, each dedicated to one of the natural diatonic chords, starting with the fundamental I, IV, and V chords and advancing to the more complex ii, iii, vi, and vii chords. Upon entering a level, learners receive text descriptions and a "Scale Circle" that introduces the target chord and its visualized musical blocks. The "Scale Circle" refers to arranging notes within the same octave in a clockwise circle by pitch class, helping learners intuitively explore the structure of the chord. The system then uses a musical composition containing the current chord along with previously learned chords to demonstrate how different harmonic structures can be built with the musical blocks, completing the composition and forming a "musical building". At the end of the level, learners reinforce their understanding by reconstructing the musical building from shuffled blocks. After finishing all the levels, learners will have acquired a comprehensive knowledge of chords and harmonic structures. If they need additional clarity on a specific chord, they can repeatedly engage with the corresponding level.

*4.3.2 Music Composition Practice.* The music composition practice flow allows learners to freely use the chords they have previously studied to create new musical works. In ArchiTone's Creation Mode, they can select a block body, tenon and mortise to form a musical block, use various musical blocks to construct harmonic structures, and ultimately complete their compositions through a combination of these harmonic structures. This flow aims at applying music theory to music composition practice, enabling users to swiftly put their knowledge into action and achieve tangible results.

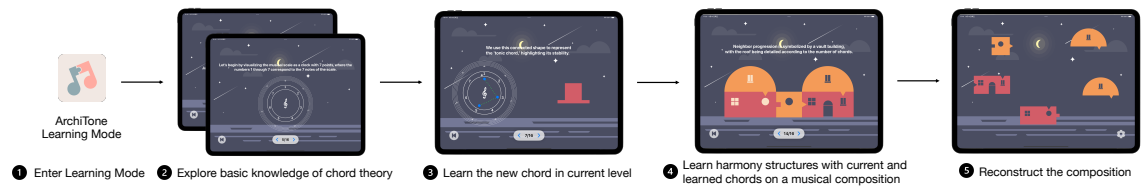

Fig. 5. Overview of the music theory learning flow. 1) Enter the learning mode of ArchiTone. 2) Explore the basic knowledge of chord theory (only appears in the first level). 3) Learn the new chord in current level. 4) Learn harmony structures using the current chord along with previously learned chords through a musical composition. 5) Let learners reconstruct the music composition by themselves.

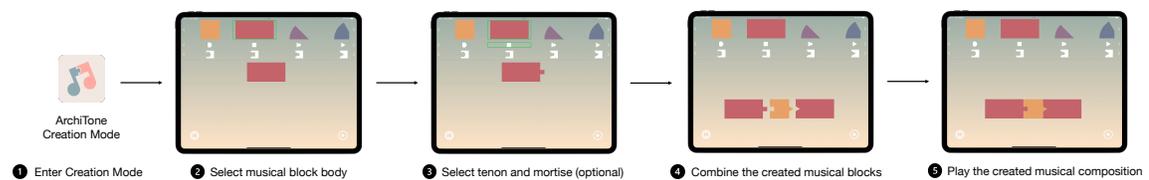

Fig. 6. Overview of the music composition practice flow. 1) Enter the creation mode of ArchiTone. 2) Select musical block body. 3) Select tenon and mortise (optional). 4) Combine the created musical blocks. 5) Play the created musical composition.

### 4.4 Implementation

ArchiTone is implemented as an iPadOS application. We utilize Figma [1] and SwiftUI [2] for user interface design and development. SwiftUI allows the rapid creation of responsive and adaptive interfaces in a declarative way. For musical blocks, we leverage the native SwiftUI struct *Shape* to draw paths instead of images for better rendering flexibility and clearer graphic quality. Additionally, the movement, attraction, and repulsion of the musical blocks are animated by *Animation* struct, creating a smoother and more cohesive overall experience.

## 5 EVALUATION

We conducted a series of user studies to thoroughly evaluate ArchiTone, focusing on the system's effectiveness in the following aspects:

- Whether users can intuitively understand music theory and its underlying connections through ArchiTone's visual design to verify the effectiveness of the visualization approach.
- Whether users can establish a connection between music theory and musical sounds to evaluate the effectiveness of interactive methods in facilitating the integration of theory and practice.
- Whether users can recognize music theory concepts from musical sounds in new contexts to assess the transferability of learning outcomes.
- Whether users find ArchiTone appealing to explore its effectiveness in enhancing interest and engagement in music education.

---

[1] https://www.figma.com
[2] https://developer.apple.com/xcode/swiftui

In the remainder of this section, we provide details on the participants and settings, the experiment procedure, and the evaluation methods.

### 5.1 Participants and Settings

This study aims to recruit adult learners who are interested in music and have experience in musical activities, to enrich the research sample with diverse musical experiences. To achieve this, we used a recruitment strategy that combined online and campus events, targeting the potential user group of ArchiTone. We focused on including learners with varied musical backgrounds to ensure that the results are relevant and valuable to a broad range of music education communities.

Participants were selected based on their passion for music and their experience with learning or self-teaching at least one musical instrument, ensuring they had a foundational understanding of music concepts. Additionally, we conducted a preliminary assessment of participants' musical abilities to confirm they had no prior knowledge of the content covered in the experiment.

A total of 14 participants were recruited and randomly assigned to either the experimental group (E) (ages 21-32; 3 females, 4 males) or the control group (C) (ages 21-32; 3 females, 4 males). The experimental group received music theory instruction through the ArchiTone system, while the control group followed traditional music theory teaching methods. Before the instructional activities began, all participants completed a pre-test to assess their baseline knowledge of music theory. After the instruction, participants took the same test again to evaluate their improvement in music theory skills.

The instructional content of this study focuses on basic harmonic theory, as harmony represents a more advanced branch of music theory that many beginners have yet to explore. Meanwhile, the material includes introductory concepts that beginners can grasp. Harmony was selected as the focus of instruction with the expectation that it will provide a solid foundation for validating the study's methods in future experimental research.

The instruction covered the definitions of the I, ii, IV, and V chords in the major scale, the construction of harmony structures, and common applications. It also covered the concepts of tonic, subdominant, and dominant chords, along with their respective roles in musical compositions. To ensure consistency and validity in comparisons, all participants—whether in the ArchiTone group or the traditional method group—were presented with the same instructional content. Both approaches also employed identical materials to explain the concepts, ensuring that differences in learning outcomes could be attributed to the teaching method rather than variations in content delivery.

### 5.2 Experiment Procedure

Before the experiment begins, researchers thoroughly explained the study's objectives, procedures, and requirements to participants. This was achieved by distributing and reviewing a detailed document that outlined the experimental framework. Participants were then asked to sign an informed consent form, ensuring they fully understand the study and their rights as participants.

To establish a baseline of participants' knowledge of music theory and aural skills, a pre-test was conducted. After the test, participants were divided into two groups: the experimental group and the control group. Both groups aimed to master the concepts of I, V, IV, and ii chords, but the learning methods differ. Participants in the experimental group first watched an introductory video for ArchiTone, an innovative music theory learning system. This video provided an overview of the app's features and capabilities. After the introduction, participants explored and self-studied using ArchiTone, with encouragement to utilize its features to deepen their understanding of music theory. Researchers

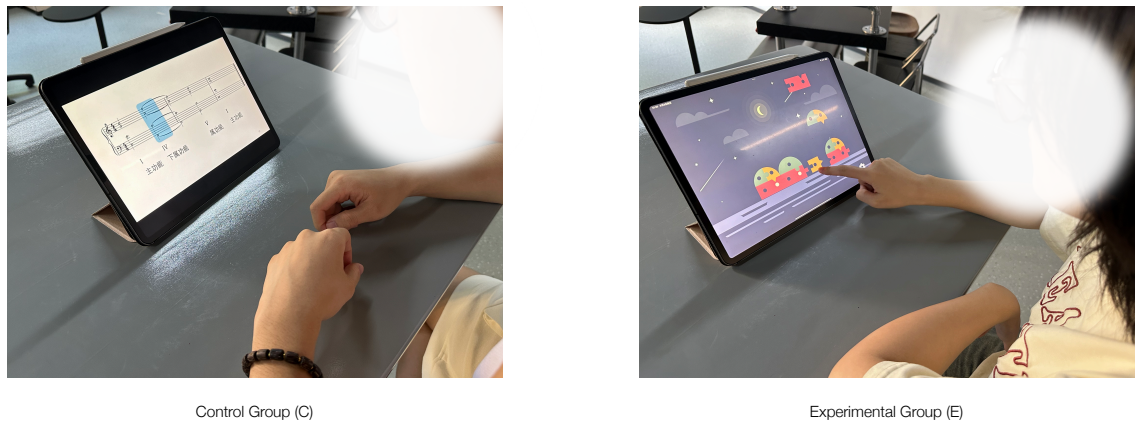

Control Group (C)                                    Experimental Group (E)

Fig. 7. Participants in the control group that watch the traditional instructional video and in the experimental group that interact with the ArchiTone installed onto an iPad for music theory learning.

were available to assist and ensured participants fully grasp the system's functionalities. In contrast, participants in the control group watched a traditional music theory instructional video covering the same chord concepts. After the video, they engaged in conventional exercises designed to reinforce their learning.

At the end of the learning phase, all participants were asked to complete the Course Experience Questionnaire (CEQ) [70] to evaluate the quality of the teaching methods they experienced. Developed by researchers from the UK and Australia, the CEQ measures key aspects of students' learning experiences [49, 50, 70] based on the concept of student learning approaches [40, 41]. It assesses how students perceive course features that promote either deep or surface learning approaches, providing insight into significant teaching aspects that influence learning, rather than measuring learning outcomes directly. In this study, we used the CEQ36 [70] version, omitting questions not relevant to our research. The CEQ employs a 5-point Likert scale, where 1 represents strong disagreement and 5 represents strong agreement. Some questions are reverse-coded; these scores will be reversed during the data analysis process for ease of interpretation.

Additionally, participants in the experimental group were also asked to complete the Post-study System Usability Questionnaire (PSSUQ) [54] to assess their satisfaction with the ArchiTone system. The PSSUQ uses a 7-point scale, where 1 indicates strong agreement and 7 indicates strong disagreement. Despite some practitioners' preference for interpreting higher scores as indicating greater satisfaction, standardized practices recommend maintaining this scale to ensure consistency unless evidence suggests it affects the item's factor structure [54]. Therefore, we adhered to the convention of interpreting lower scores as indicative of higher satisfaction. To evaluate the effectiveness of the learning methods, participants took the same test that was administered before the experiment. This allowed for a direct comparison of their music theory knowledge and aural skills before and after the intervention. Following the experiment, we conducted feedback interviews with all 14 participants from both the experimental and control groups to gather insights into their experiences and opinions regarding ArchiTone and the traditional methods.

### 5.3 Assessment

*5.3.1 Music Theory and Aural Skills.* The test for Music theory and aural skills is designed to evaluate changes in participants' cognitive levels before and after music theory instruction. Its content is aligned with the curriculum to ensure specificity and effectiveness. The test is structured into three main modules, each designed to assess participants' music theory skills at different levels:

- **Concept Understanding Ability:** This module evaluates participants' comprehension of fundamental music theory concepts, including chords, chord progressions and harmony structures.
- **Aural Discrimination Ability:** This module evaluates participants' proficiency in recognizing and analyzing chords in musical pieces through aural tests. It measures their accuracy in identifying music theory concepts by ear.
- **Concept Application Ability:** This module evaluates participants' capacity to apply theoretical concepts in analyzing real musical works. It assesses their proficiency in recognizing and utilizing learned knowledge within complex musical contexts.

To ensure a comprehensive and accurate assessment of participants' progress in music theory learning, this study invited professional music theory educators to design a set of questions based on the three key abilities outlined. This multidimensional approach allows for a thorough evaluation of how different teaching methods impact learning outcomes. Each module includes three multiple-choice questions. Researchers will compare participants' scores before and after using ArchiTone or the traditional method to assess learning effectiveness. This comparative analysis will provide empirical evidence on the efficacy of the different teaching methods.

*5.3.2 Teaching Experience.* To evaluate participants' experiences during the teaching phase, the assessment incorporates six indicators from the CEQ questionnaire:

- **Appropriate Assessment (AA):** This factor reflects whether the assessment methods used in the course are fair, effective, and accurately reflect students' learning outcomes. It includes the design and implementation of exams, assignments and other evaluation tools.
- **Appropriate Workload (AW):** This factor reflects whether the course workload is reasonable and matches students' learning capacity and time constraints.
- **Clear Goals and Standards (CG):** This factor reflects the clarity of course objectives and learning expectations. Clear goals and standards are crucial for helping students understand course requirements, the outcomes they need to achieve, and how to measure their own progress.
- **Generic Skills (GS):** This factor reflects whether the course supports the development of interdisciplinary or cross-domain skills, such as critical thinking, problem-solving, self-directed learning, etc.
- **Good Teaching (GT):** This factor reflects the performance of the teaching entity (ArchiTone or the instructor of the video) throughout the instructional process. It assesses teaching methods, interaction, support for students' learning, and the instructor's effectiveness in explaining and conveying course content.
- **Emphasis on Independence (IN):** This factor reflects whether the course promotes the development of independent learning skills, including self-directed planning and self-assessment.
- **Overall:** An additional indicator to evaluate participants' overall satisfaction with the teaching.

In the experiment, we compare the teaching experiences of participants in the experimental group with those in the control group to evaluate the effectiveness of the two teaching methods.

*5.3.3 Teaching System Usability.* Since ArchiTone is a system working on mobile devices, we used the PSSUQ to evaluate participants' perceived satisfaction with the system. The PSSUQ provides an overall score and assesses three specific dimensions: System Quality (SysQual), Information Quality (InfoQual), and Interface Quality (IntQual).

## 6 RESULTS

All participants completed the all experimental procedures successfully. We conduct a comprehensive analysis of the experimental data, including participants' pre- and post-test scores, feedback from the CEQ, and responses from the PSSUQ.

### 6.1 Pre- and Post-Test for Music Theory and Aural Skills

Table 1 presents the accuracy rates of participants in both the pre-test and post-test, including the mean accuracy rates ($M$) and standard deviation ($SD$) for the concept understanding, aural discrimination and concept application modules, respectively. To assess differences between the experimental and control groups, independent samples t-tests are conducted for each module and overall. In the pre-test, no significant differences are found between the groups in terms of concept understanding ($p = 0.477$), aural discrimination ($p = 0.740$), concept application ($p = 0.232$), and overall ($p = 0.452$). These results indicate that both groups had comparable levels of musical ability at the start.

In the post-test, all participants achieve perfect scores in the theoretical concept understanding module. This result suggests that our visualization method effectively supports participants' comprehension of music theory concepts. The control group also attains perfect scores in this module, indicating the high quality of the instructional videos. While this result does not explicitly show the advantages of our visualization design, we plan to investigate this further through follow-up modules. Compared to their performance in concept understanding, the experimental group demonstrates significant advantages in aural discrimination and concept application. Specifically, the experimental group achieves accuracy scores of 95.2% ($SD = 0.117$) and 81.0% ($SD = 0.243$) in the aural discrimination and concept application modules, while the control group achieves accuracy rates of 61.9% ($SD = 0.278$) and 28.6% ($SD = 0.330$), respectively. These results highlight a substantial advantage for the experimental group and support our hypothesis. In terms of overall accuracy rate, the experimental group also outperforms the control group, with a significant difference between the two groups as revealed by the t-test ($p = 0.003$).

Table 1. The results of participants in both the pre- and post-tests for music theory and aural skills. E refers to the experimental group and C refers to the control group.

| Test | Group | Concept Understanding | | | Aural Discrimination | | | Concept Application | | | Overall | | |
|---|---|---|---|---|---|---|---|---|---|---|---|---|---|
| | | $M$ | $SD$ | $p$ | $M$ | $SD$ | $p$ | $M$ | $SD$ | $p$ | $M$ | $SD$ | $p$ |
| **Pre** | E | 0.381 | 0.330 | 0.476 | 0.190 | 0.243 | 0.740 | 0.238 | 0.233 | 0.231 | 0.254 | 0.203 | 0.452 |
| | C | 0.238 | 0.343 | | 0.143 | 0.243 | | 0.095 | 0.151 | | 0.159 | 0.221 | |
| **Post** | E | 1.000 | 0.000 | – | 0.952 | 0.117 | 0.019* | 0.810 | 0.243 | 0.009** | 0.921 | 0.115 | 0.003** |
| | C | 1.000 | 0.000 | | 0.619 | 0.278 | | 0.286 | 0.330 | | 0.619 | 0.167 | |

*: $p<0.05$, **: $p<0.01$, ***: $p<0.001$

### 6.2 CEQ for Teaching Experience

The CEQ is used in this study to assess participants' experiences with the teaching sessions. As shown in Table 2, there are no significant differences between the control and experimental groups across most evaluation metrics, including AA, AW, CG, GT and IN. These results align with our expectations and suggest the fairness of the experimental design. If there are significant differences between these key metrics, the observed changes in pre- and post-test results might not be fully attributable to the effects of the visualization and interaction designs. For the GS metric, the experimental group ($M = 4.400, SD = 0.490$) performs better than the control group ($M = 3.867, SD = 0.806$) with a significant difference ($p = 0.043$). It indicates that participants in the experimental group feel more confident in the skills they gained through the ArchiTone, which is also reflected in the post-test results. Furthermore, the experimental group ($M = 4.600, SD = 0.490$) also outperforms the control group ($M = 3.400, SD = 0.800$) in the overall evaluation, due to the gamified design in ArchiTone that enhances the overall learning experience, surpassing traditional teaching methods.

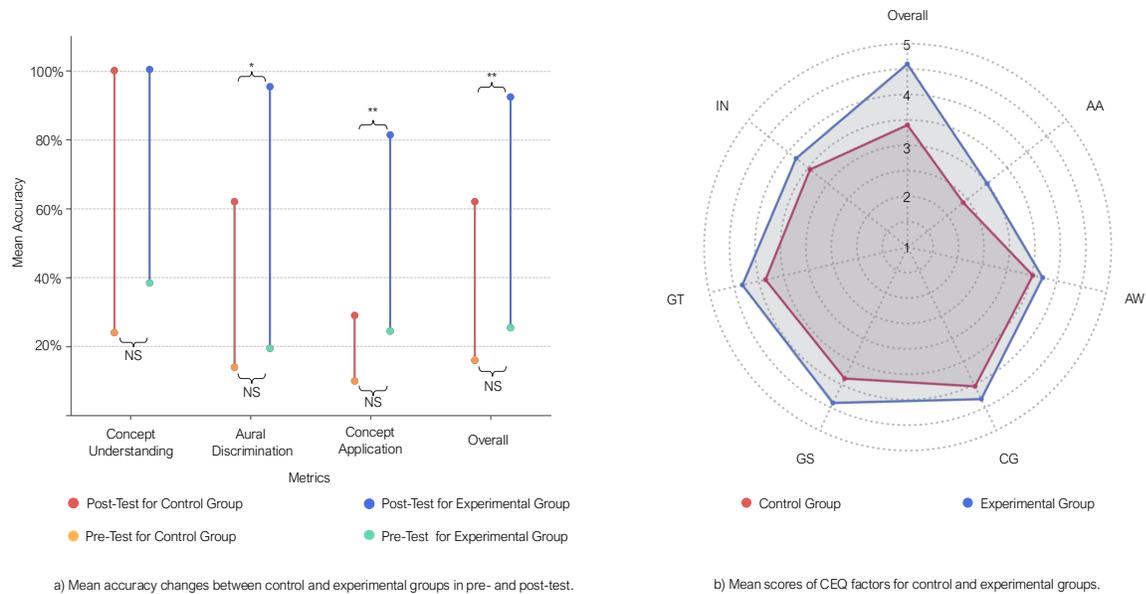

a) Mean accuracy changes between control and experimental groups in pre- and post-test.

b) Mean scores of CEQ factors for control and experimental groups.

Fig. 8. Comparison of mean accuracy changes and CEQ factors between control and experimental groups. a) Mean accuracy changes in the pre- and post-test for music theory and aural skills between control and experimental groups. b) Mean scores of CEQ factors for control and experimental groups: AA, AW, CG, GS, GT, In and Overall.

### 6.3 PSSUQ For Teaching System Usability

We also utilize the PSSUQ to assess system's performance across several key dimensions, including efficiency, ease of use, interface appeal, and overall user satisfaction. As detailed in Table 3, we calculate scores for the three factors: SysQual, InfoQual, and IntQual, with the percentage distribution of each factor illustrated in Fig. 9. The analysis consistently reveals positive evaluations across all factors, with average scores reflecting strong user satisfaction. These findings suggest that the system is not only easy to use but also effectively meets user expectations. Specifically, the high score for SysQual indicates that users find the system easy to learn and use, without facing a steep learning curve. The high score for InfoQual reflects the effectiveness of our visualization design, ensuring clarity and ease of understanding.

Table 2. The results of CEQ for teaching experience. E refers to the experimental group and C refers to the control group. AA refers to Appropriate Assessment, AW refers to Appropriate Workload, CG refers to Clear Goals and Standards, GS refers to Generic Skills, GT refers to Good Teaching and IN refers to Emphasis on Independence.

| Factor | E | | | C | | | $p$ |
|---|---|---|---|---|---|---|---|
| | $M$ | $SD$ | Cronbach−$\alpha$ | $M$ | $SD$ | Cronbach−$\alpha$ | |
| **AA** | 2.400 | 0.917 | | 3.000 | 1.483 | | 0.316 |
| **AW** | 3.520 | 0.943 | | 3.720 | 1.001 | | 0.480 |
| **CG** | 4.040 | 0.916 | | 4.320 | 0.733 | | 0.248 |
| **GS** | 3.867 | 0.806 | 0.895 | 4.400 | 0.490 | 0.842 | 0.043* |
| **GT** | 3.867 | 1.024 | | 4.333 | 0.596 | | 0.152 |
| **IN** | 3.450 | 1.071 | | 3.800 | 1.077 | | 0.322 |
| **Overall** | 3.400 | 0.800 | | 4.600 | 0.490 | | 0.034* |

*: $p<0.05$, **: $p<0.01$, ***: $p<0.001$

The high score for IntQual further confirms that the system not only fulfills users' functional needs but also delivers a comfortable and user-friendly experience.

Table 3. The results of PSSUQ for teaching system usability. SysQual refers to System Quality, InfoQual refers to Information Quality, and IntQual refers to Interface Quality. Lower PSSUQ scores indicate higher satisfaction.

| Factor | $M$ | $SD$ | Cronbach − $\alpha$ |
|---|---|---|---|
| **SysQual** | 1.933 | 0.928 | |
| **InfoQual** | 1.611 | 0.756 | 0.916 |
| **IntQual** | 1.333 | 0.471 | |
| **Overall** | 1.600 | 0.707 | |

In summary, the questionnaire results underscore the positive role of ArchiTone's design in enhancing music theory learning, with no negative effects on the learning process. Moreover, the high Cronbach's alpha ($\alpha$ = 0.916) confirms the reliability of the measurements, demonstrating strong internal consistency in user feedback.

# 7 DISCUSSION

## 7.1 Interactive Visualization for Abstract Knowledge Construction

For most learners, music education presents significant challenges, largely due to the difficulty in grasping abstract concepts and their interconnections within music theory. Furthermore, the abstractness of music theory often hinders learners from effectively linking theoretical knowledge with real musical sounds. ArchiTone addresses these challenges through a series of meticulously designed strategies to enhance learners' comprehension of music theory.

*7.1.1 Making Abstract Concepts Visual and Interactive.* Visualization is an effective strategy for learning abstract music theory [25, 39]. By converting abstract concepts into visual and interactive forms, learners gain a more intuitive and engaging understanding of theoretical principles. Our pre- and post-tests revealed significant improvements in theoretical accuracy, confirming the feasibility and effectiveness of this approach. Learners with similar starting proficiency achieved 100% accuracy using both ArchiTone and traditional methods, indicating that ArchiTone's interactive visualizations are equally effective in enhancing comprehension. However, user studies show that while both methods improve accuracy, learners using ArchiTone find it easier to grasp complex concepts and experience reduced cognitive load. In contrast, traditional methods often lead to fatigue on visual and auditory. Additionally, learners using traditional methods frequently need to rewind and rewatch instructional videos, a problem less common with ArchiTone. These findings highlight the critical importance of visualization in mastering abstract music theory.

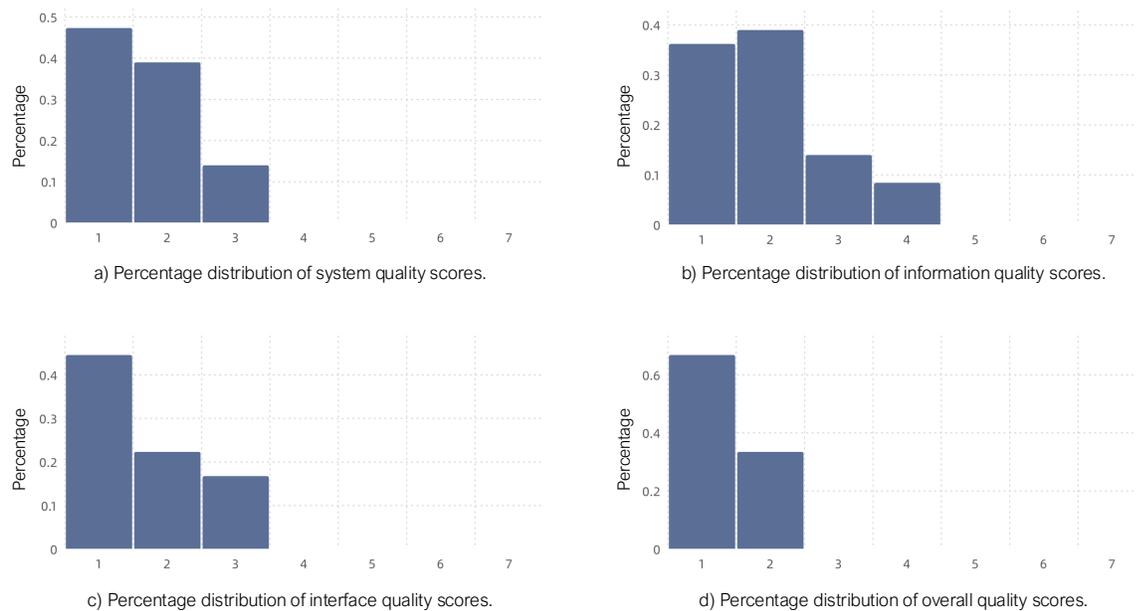

Fig. 9. The percentage distribution of PSSUQ factor scores: a) System Quality, b) Information Quality, c) Interface Quality, and d) Overall Quality. Lower PSSUQ scores indicate higher satisfaction.

*7.1.2 Promoting Integration and Application of Knowledge through Analogies.* In addition, we find that analogies can help learners transfer previously acquired knowledge to new contexts, facilitating knowledge integration and application. By linking new concepts with existing ones, analogies not only deepen understanding but also improve problem-solving skills, enabling learners to apply their knowledge more flexibly in real-world situations. In our experiment, participants demonstrate significant differences in performance on tests of acquired knowledge and tests requiring knowledge transfer after using ArchiTone and the traditional method. This suggests that ArchiTone effectively addresses the limitations of traditional methods in promoting knowledge integration and application, enhancing learners' adaptability.

## 7.2 Gamification for Enhanced Engagement and Immersion

Our formative investigation indicates that after learners grasp abstract music theory, repetitive practice is still essential to solidify their understanding. This poses a second challenge, as the monotony of practice often leads to negative emotions, reducing engagement and ultimately causing learners to give up [20, 29, 62]. To counter this negative cycle, educators must design interactive approaches that actively engage learners and sustain their interest.

*7.2.1 Designing Gamified Interface and Interaction.* Gamification is increasingly recognized as an effective educational approach, capable of balancing enjoyment with educational objectives [30, 31]. In ArchiTone's design, gamification serves not only as a motivational tool but also as a core mechanism driving the learning process. Our user studies show that the integration of a gamified interface creates a highly interactive learning environment, enabling learners to naturally focus on key instructional content. By establishing clear task objectives, ArchiTone enhances learners' understanding of learning goals, thereby improving overall learning efficiency. Additionally, we find that gamified designs effectively guide participants' visual focus through elements such as color and shape, ensuring attention remains centered on key points. In contrast, traditional music education typically follows a linear way, where learners progress through standardized steps. This method often struggles to maintain engagement, leading to disinterest or confusion, making it difficult for learners to sustain high levels of engagement over extended periods.

*7.2.2 Providing Real-Time Feedback During Interaction.* Real-time feedback during interaction allows learners to immediately assess the accuracy of their actions or decisions, enabling them to quickly adjust and refine their learning strategies [5, 13, 64]. Our experiments demonstrate that immediate error correction not only reduces the frustration caused by cumulative mistakes but also significantly boosts learners' confidence, encouraging more active participation in learning activities. For example, in ArchiTone, when the tenon and mortise structures of two blocks align correctly, a "click" sound confirms a successful attraction, while a mismatch causes the blocks to repel. This feedback mechanism helps learners instantly recognize the accuracy of their actions, allowing them to adjust strategies and try further. Some participants found this feedback so engaging that they explored all possible block combinations. By providing immediate and enjoyable feedback, learners become more immersed in the learning process, which continuously reinforces their understanding and enhances knowledge retention.

## 7.3 Constructivism for Lifelong Education and Practice

In the context of the increasing focus on lifelong music education [66], constructivism offers a robust approach for supporting ongoing personal learning and adaptation. Inspired by constructivism and the LEGO education system, ArchiTone develops a unique system aimed at facilitating lifelong education and practice.

*7.3.1 Implementing Progressive Learning Level by Level.* Progressive learning provides a framework for lifelong education by systematically building knowledge and skills, thereby supporting individuals in adapting to changes and pursuing new goals. ArchiTone employs a level-based system where learners gain new knowledge while applying previously acquired skills at each level. User studies indicate that ArchiTone is more effective than traditional teaching methods in developing essential transferable skills, such as adaptability and critical thinking. These skills help learners explore new fields, address complex problems, and achieve a competitive edge in a dynamic environment. Furthermore, our findings show that participants using ArchiTone have a shorter average response time ($M = 184$ seconds) on the post-test of music theory compared to those using traditional methods ($M = 229$ seconds). This suggests that ArchiTone promotes more efficient knowledge acquisition and delivers superior teaching quality.

*7.3.2 Combining Theory with Practice.* Combining theory with practice is crucial in music education, as it assesses learners' ability to apply knowledge to new contexts. Objective experimental data reveal that ArchiTone significantly outperforms traditional teaching methods in enhancing knowledge transfer, especially in the generalization accuracy sections of music theory tests, showing substantial progress among participants. Additionally, usability surveys indicate high ratings for ArchiTone's system quality, information quality, and interface quality. In the creative mode that allows free practice, most participants successfully produced compositions that were both sound beautiful and adhered to music theory and showed strong interest in the concepts of musical "blocks" and "buildings". Some participants developed musical architectures not covered in the learning mode, further validating that the integration of theory and practice not only enhances learners' hands-on skills but also deepens their understanding and application of complex musical concepts.

### 7.4 Limitations and Future Work

Firstly, as noted by some participants in our empirical study, the learning mode of our system relies heavily on text, which can lead to distractions and negatively affect comprehension and retention, thereby reducing learning efficiency. Although we have made efforts to simplify the text content, this approach does not fully address the issue, and excessive simplification may result in unclear information, impairing learning effectiveness. A potential solution is to integrate voice assistance, such as audio narration, to complement the text. Additionally, we are exploring the future implementation of AI-based personalization, which would dynamically adjust both text and narration to better meet the needs and progress of individual learners.

Secondly, ArchiTone provides only a single musical work for practice at each level currently, which may restrict learners' exposure to various musical styles and structures. To overcome this limitation, we plan to develop an automated music segmentation tool. This tool will allow users to input a complete musical work, analyze its chord and harmony structures, and generate corresponding musical blocks. This enhancement will offer a broader range of practice materials and enrich the learning experience.

Thirdly, our evaluation of ArchiTone has been limited to short-term music education. While this evaluation provides valuable insights into the system's effectiveness, it may not fully reflect its long-term impact and users' sustained engagement. Future work will focus on long-term follow-up studies to examine the system's effectiveness over extended periods and its influence on learners across different stages of music education.

Fourth, the participants in this study were all well-educated young adults, and the sample size was relatively small. As a result, the findings may not be broadly representative. Future research should involve a larger and more diverse group of users to validate the results further and improve their universality.

## 8 CONCLUSION

We presented ArchiTone, a LEGO-inspired gamified system that applies the principles of constructivism for visualized music education. We discussed 1) our formative investigation and its resulting design implications, 2) music visualization and interaction design involved in the creation of this system, and 3) the results of a multi-dimensional evaluation study that demonstrated ArchiTone's efficacy in helping learners understand abstract music theory and apply it to music praxis. Specifically, ArchiTone effectively introduced music theory concepts (chords and harmony structures) by visualizing them as musical blocks and buildings, established a connection between music theory and musical sounds for learners, and supported them in applying the acquired knowledge to music recognition and composition with a more engaging and enhanced experience. Having identified improvements to the system, including future work on designing a voice-assisted system and conducting long-term tests with more diverse groups, our contributions not only address current limitations but also pave the way for future research in integrating visualization, gamification and constructivism into educational tools.

## REFERENCES

[1] Bahjat Altakhayneh. 2020. The Impact of Using the LEGO Education Program on Mathematics Achievement of Different Levels of Elementary Students. *European Journal of Educational Research* 9, 2 (2020), 603–610.
[2] Roya Jafari Amineh and Hanieh Davatgari Asl. 2015. Review of constructivism and social constructivism. *Journal of social sciences, literature and languages* 1, 1 (2015), 9–16.

[3] Brian J Arnold. 2014. Gamification in education. *Proceedings of the American society of Business and Behavioral Sciences* 21, 1 (2014), 32–39.
[4] Fatimah Zahra Ros Azman and Muhamad Fairus Kamaruzaman. 2016. Integration of traditional music through mobile game in inspiring Malaysian youths' enthusiasm. In *2016 IEEE 8th International Conference on Engineering Education (ICEED)*. IEEE, 115–119.
[5] Roghayeh Barmaki and Charles E Hughes. 2015. Providing real-time feedback for student teachers in a virtual rehearsal environment. In *Proceedings of the 2015 ACM on International Conference on Multimodal Interaction*. 531–537.
[6] Michele Biasutti and Eleonora Concina. 2013. MUSIC EDUCATION AND TRANSFER OF LEARNING. *Journal of Communications Research* 5, 3 (2013).
[7] Karen Burland. 2021. Ensemble participation and personal development. *Together in Music: Coordination, Expression, Participation* 218 (2021).
[8] Gabriel Dias Cantareira, Luis Gustavo Nonato, and Fernando V Paulovich. 2016. Moshviz: A detail+ overview approach to visualize music elements. *IEEE Transactions on Multimedia* 18, 11 (2016), 2238–2246.
[9] Ilaria Caponetto, Jeffrey Earp, and Michela Ott. 2014. Gamification and education: A literature review. In *European conference on games based learning*, Vol. 1. Academic Conferences International Limited, 50.
[10] Chi Wai Jason Chen. 2015. Mobile learning: Using application Auralbook to learn aural skills. *International journal of music education* 33, 2 (2015), 244–259.
[11] Terence Cook, Ashlin RK Roy, and Keith M Welker. 2019. Music as an emotion regulation strategy: An examination of genres of music and their roles in emotion regulation. *Psychology of Music* 47, 1 (2019), 144–154.
[12] Nataša Cvijanović and Danica Mojić. 2018. LEGO material in the programme of early childhood and preschool education. *Croatian Journal of Education: Hrvatski časopis za odgoj i obrazovanje* 20 (2018), 25–45.
[13] Galina Deeva, Daria Bogdanova, Estefanía Serral, Monique Snoeck, and Jochen De Weerdt. 2021. A review of automated feedback systems for learners: Classification framework, challenges and opportunities. *Computers & Education* 162 (2021), 104094.
[14] Guillaume Denis and Pierre Jouvelot. 2004. Building the case for video games in music education. In *Second International Computer Game and Technology Workshop*. Citeseer, 156–161.
[15] Guillaume Denis and Pierre Jouvelot. 2005. Motivation-driven educational game design: applying best practices to music education. In *Proceedings of the 2005 ACM SIGCHI International Conference on Advances in computer entertainment technology*. 462–465.
[16] Morwaread M Farbood, Egon Pasztor, and Kevin Jennings. 2004. Hyperscore: a graphical sketchpad for novice composers. *IEEE Computer Graphics and Applications* 24, 1 (2004), 50–54.
[17] Eileen Ferrer, Polong Lew, Sarah Jung, Emilia Janeke, Michelle Garcia, Cindy Peng, George Poon, Vinisha Rathod, Sharon Beckwith, and Chick Tam. 2014. Playing music to relieve stress in a college classroom environment. *College Student Journal* 48, 3 (2014), 481–494.
[18] Edwin Gordon. 2007. *Learning sequences in music: A contemporary music learning theory*. Gia Publications.
[19] Richard F Grunow. 2005. Music learning theory: A catalyst for change in beginning instrumental music instruction. *The development and practical application of music learning theory* (2005), 179–200.
[20] Laurent Guirard. 1999. Abandonner la musique?: psychologie de la motivation et apprentissage musical. (1999).
[21] Hua Guo. 2018. Application of a Computer-Assisted Instruction System Based on Constructivism. *International Journal of Emerging Technologies in Learning* 13, 4 (2018).
[22] Julian Hernandez-Serrano, Ikseon Choi, and David H Jonassen. 2000. Integrating constructivism and learning technologies. *Integrated and holistic perspectives on learning, instruction and technology: Understanding complexity* (2000), 103–128.
[23] Lenita Hietanen and Heikki Ruismäki. 2017. The use of a blended learning environment by primary school student teachers to study music theory. *The European Journal of Social & Behavioural Sciences* 19, 2 (2017), 2393–2404.
[24] Rumi Hiraga, Reiko Mizaki, and Issei Fujishiro. 2002. Performance visualization: a new challenge to music through visualization. In *Proceedings of the tenth ACM international conference on Multimedia*. 239–242.
[25] Rumi Hiraga, Fumiko Watanabe, and Issei Fujishiro. 2002. Music learning through visualization. In *Second International Conference on Web Delivering of Music, 2002. WEDELMUSIC 2002. Proceedings*. IEEE, 101–108.
[26] Esa-Matti Jarvinen. 1998. The Lego/Logo Learning Environment in Technology Education: An Experiment in a Finnish Context. *Journal of technology education* 9, 2 (1998), 47–59.
[27] Elissa Johnson-Green. 2018. Musical architects: Immersive learning through design thinking in a kindergarten music composition curriculum. *Visions of Research in Music Education* 31, 1 (2018), 2.
[28] David H Jonassen. 1994. Technology as cognitive tools: Learners as designers. *ITForum Paper* 1, 1 (1994), 67–80.
[29] DT Kenny. 2009. Negative emotions in music making: Performance anxiety. *Handbook of music and emotion: Theory, research, applications* (2009), 425–451.
[30] Christoph Klimmt. 2003. Dimensions and determinants of the enjoyment of playing digital games: A three-level model. In *Level up: Digital games research conference*. 246–257.
[31] Raph Koster. 2013. *Theory of fun for game design*. " O'Reilly Media, Inc.".
[32] Steven Geoffrey Laitz. 2012. The complete musician: An integrated approach to tonal theory, analysis, and listening. (2012).
[33] Jaejun Lee, Hyeyoon Cho, and Yonghyun Kim. 2023. Bean Academy: A Music Composition Game for Beginners with Vocal Query Transcription. In *Extended Abstracts of the 2023 CHI Conference on Human Factors in Computing Systems*. 1–6.

[34] Yu Liang and Martijn C Willemsen. 2021. Interactive music genre exploration with visualization and mood control. In *Proceedings of the 26th International Conference on Intelligent User Interfaces*. 175–185.

[35] Hugo B Lima, Carlos GR Dos Santos, and Bianchi S Meiguins. 2021. A survey of music visualization techniques. *ACM Computing Surveys (CSUR)* 54, 7 (2021), 1–29.

[36] Jörgen Lindh and Thomas Holgersson. 2007. Does lego training stimulate pupils' ability to solve logical problems? *Computers & education* 49, 4 (2007), 1097–1111.

[37] Yan Liu, Hongbing Liu, Yan Xu, and Hongying Lu. 2020. Online English reading instruction in the ESL classroom based on constructivism. *Journal of Educational Technology Systems* 48, 4 (2020), 539–552.

[38] Yi Lu, Xiaoye Wang, Jiangtao Gong, and Yun Liang. 2022. ChordAR: an educational AR game design for children's music theory learning. *Wireless communications and mobile computing* 2022, 1 (2022), 5268586.

[39] Delfina Malandrino, Donato Pirozzi, and Rocco Zaccagnino. 2018. Visualization and music harmony: Design, implementation, and evaluation. In *2018 22nd International Conference Information Visualisation (IV)*. IEEE, 498–503.

[40] F Marton and S Booth. 1997. Learning and Awareness. New Jersey: Lawrence Earlbum Associates. *Inc., Publishers* (1997).

[41] Ference Marton and Roger Säljö. 1976. On qualitative differences in learning: I—Outcome and process. *British journal of educational psychology* 46, 1 (1976), 4–11.

[42] Saif Mohammed and László Kinyó. 2020. Constructivist theory as a foundation for the utilization of digital technology in the lifelong learning process. *Turkish Online Journal of Distance Education* 21, 4 (2020), 90–109.

[43] Jose O Montesa-Andres, Fernando J Garrigós-Simón, and Yeamduan Narangajavana. 2014. A proposal for using Lego serious play in education. *Innovation and teaching technologies: New directions in research, practice and policy* (2014), 99–107.

[44] Kimberly Sena Moore. 2013. A systematic review on the neural effects of music on emotion regulation: Implications for music therapy practice. *Journal of music therapy* 50, 3 (2013), 198–242.

[45] HyunJu Park, Soo-yong Byun, Jaeho Sim, Hye-Sook Han, and Yoon Su Baek. 2016. Teachers' perceptions and practices of STEAM education in South Korea. *Eurasia Journal of Mathematics, Science and Technology Education* 12, 7 (2016), 1739–1753.

[46] Matevž Pesek, Žiga Vučko, Peter Šavli, Alenka Kavčič, and Matija Marolt. 2020. Troubadour: A gamified e-learning platform for ear training. *IEEE Access* 8 (2020), 97090–97102.

[47] Giulio Pitteri, Edoardo Micheloni, Carlo Fantozzi, and Nicola Orio. 2021. Listen by Looking: A framework to support the development of serious games for live music. *Entertainment Computing* 37 (2021), 100394.

[48] Stephanie E Pitts. 2017. What is music education for? Understanding and fostering routes into lifelong musical engagement. *Music Education Research* 19, 2 (2017), 160–168.

[49] Paul Ramsden. 1991. A performance indicator of teaching quality in higher education: The Course Experience Questionnaire. *Studies in higher education* 16, 2 (1991), 129–150.

[50] P Ramsden. 2003. Learning to teach in higher education.

[51] Thomas A Regelski. 2005. Music and Music Education: Theory and praxis for 'making a difference'. *Educational Philosophy and Theory* 37, 1 (2005), 7–27.

[52] Grega Repovš and Alan Baddeley. 2006. The multi-component model of working memory: Explorations in experimental cognitive psychology. *Neuroscience* 139, 1 (2006), 5–21.

[53] Heikki Ruismäki and Antti Juvonen. 2006. The good, the bad and the ugly memories from the school art subjects education: the teaching of art subjects in narratives of kindergarten teacher students. 1–13.

[54] Jeff Sauro and James R Lewis. 2016. *Quantifying the user experience: Practical statistics for user research*. Morgan Kaufmann.

[55] Amer Mutrik Sayaf. 2023. Adoption of E-learning systems: An integration of ISSM and constructivism theories in higher education. *Heliyon* 9, 2 (2023).

[56] Viktor Shurygin, Roza Ryskaliyeva, Elena Dolzhich, Svetlana Dmitrichenkova, and Alexander Ilyin. 2022. Transformation of teacher training in a rapidly evolving digital environment. *Education and Information Technologies* (2022), 1–20.

[57] Peti Simon and Tamas Szabo. 2013. *Music: social impacts, health benefits and perspectives*. Nova Publishers.

[58] Sean Soraghan, Felix Faire, Alain Renaud, and Ben Supper. 2018. A new timbre visualization technique based on semantic descriptors. *Computer Music Journal* 42, 1 (2018), 23–36.

[59] Isabelle ML Souza, Wilkerson L Andrade, Lívia MR Sampaio, and Ana Liz Souto O Araujo. 2018. A Systematic Review on the use of LEGO® Robotics in Education. In *2018 IEEE frontiers in education conference (FIE)*. IEEE, 1–9.

[60] J. Sweller. 1999. *Instructional Design in Technical Areas*. ACER Press.

[61] Michael Taenzer, Burkhard C Wünsche, and Stefan Müller. 2019. Analysis and visualisation of music. In *2019 International Conference on Electronics, Information, and Communication (ICEIC)*. IEEE, 1–6.

[62] Töres Theorell and Fredrik Ullén. 2018. Music practice and emotion handling. *Music and Public Health: A Nordic Perspective* (2018), 55–67.

[63] Myriam V Thoma, Stefan Ryf, Changiz Mohiyeddini, Ulrike Ehlert, and Urs M Nater. 2012. Emotion regulation through listening to music in everyday situations. *Cognition & emotion* 26, 3 (2012), 550–560.

[64] Mike Tissenbaum and Jim Slotta. 2019. Supporting classroom orchestration with real-time feedback: A role for teacher dashboards and real-time agents. *International Journal of Computer-Supported Collaborative Learning* 14 (2019), 325–351.

[65] Ali Korkut Uludag and Ugur Kartal Satir. 2023. Seeking alternatives in music education: The effects of mobile technologies on students' achievement in basic music theory. *International Journal of Music Education* (2023), 02557614231196972.
[66] Kari K Veblen. 2018. Adult music learning in formal, nonformal, and informal contexts. *Special needs, community music, and adult learning: An Oxford handbook of music education* 4 (2018), 243–256.
[67] Graham F Welch, Michele Biasutti, Jennifer MacRitchie, Gary E McPherson, and Evangelos Himonides. 2020. The impact of music on human development and well-being. , 1246 pages.
[68] Graham F Welch, Evangelos Himonides, Jo Saunders, Ioulia Papageorgi, and Marc Sarazin. 2014. Singing and social inclusion. *Frontiers in psychology* 5 (2014), 803.
[69] Heidi Westerlund and Lauri Väkevä. 2011. Who needs theory anyway? The relationship between theory and practice of music education in a philosophical outlook. *British Journal of Music Education* 28, 01 (2011), 37–49.
[70] Keithia L Wilson, Alf Lizzio, and Paul Ramsden. 1997. The development, validation and application of the Course Experience Questionnaire. *Studies in higher education* 22, 1 (1997), 33–53.
[71] Nechama Yehuda. 2011. Music and stress. *Journal of Adult Development* 18 (2011), 85–94.
[72] Beste F Yuksel, Kurt B Oleson, Lane Harrison, Evan M Peck, Daniel Afergan, Remco Chang, and Robert JK Jacob. 2016. Learn piano with BACh: An adaptive learning interface that adjusts task difficulty based on brain state. In *Proceedings of the 2016 CHI conference on human factors in computing systems*. 5372–5384.
[73] Yinsheng Zhou, Graham Percival, Xinxi Wang, Ye Wang, and Shengdong Zhao. 2010. Mogclass: a collaborative system of mobile devices forclassroom music education. In *Proceedings of the 18th ACM international conference on Multimedia*. 671–674.